 \documentclass[final,3p,times]{elsarticle}

\usepackage{latexsym}
\usepackage{amssymb,latexsym} 
\usepackage[fleqn]{amsmath}
\usepackage{graphicx}
\usepackage{dcolumn}
\usepackage{bm}
\usepackage{caption}
\usepackage{subcaption}
\usepackage{grffile}
\usepackage{multirow}
\usepackage[colorlinks=true,allcolors=blue]{hyperref}
\usepackage[mathlines]{lineno}

\newcommand{\irow}[1]{
  \begin{matrix}(\,#1\,)\end{matrix}}

\usepackage{amsthm}

\journal{Commun Nonlinear Sci Numer Simulat}
\begin{document}

\begin{frontmatter}



\title{Frequency locking and travelling burst sequences in community structured network of inhibitory neurons with differing time-scales}

 \author[label1]{Kunal Mozumdar}
 \author[label1]{G. Ambika}
 \address[label1]{Indian Institute of Science Education and Research, Pune, India - 411008 }



\begin{abstract}
We report the emergent dynamics of a community structured modular network of chaotic Hindmarsh-Rose (HR) neurons with inhibitory synapses. We find the inhibitory coupling between the neuronal modules lead to complete synchronization of neurons in a module, and also pushes modules into interesting sequences of travelling burst patterns. When dynamical time-scales vary for neurons in different modules, hence breaking the symmetry among them, we see specific sequences of travelling burst patterns that are characteristic of the time-scale mismatch and coupling strengths. Thus for a modular network with two time-scales, the neuronal communities enter into synchronized frequency locked clusters with the bursting sequences having recurring patterns. Our study provides a complete characterization of the spatio-temporal regularity in terms of frequency locking for temporal order and burst sequence patterns for spatial order in the collective dynamics of the neuronal clusters. Our results have significance in the process of information coding in terms of frequency of firing dynamics among neurons and in selective communication based on the sequences of bursts.
\end{abstract}


\end{frontmatter}


\section{Introduction}
\label{sec-1}

Studies related to the brain and its various structural and dynamical features have been a very interesting and challenging area of research in recent times. This is mainly because the brain is highly complex with over $10^9$ neurons connected via complex networks perfected by evolution. Although reproducing the dynamics of the entire brain is close to impossible, due to its intricate connectome and constituent levels of its complexity, there are several generalizations in the neuronal architecture and dynamical features of the brain. These generalizations can be abstracted to give very useful mathematical models of neuronal networks, which can be extended to model the real structures in the brain. \cite{bullmore2009complex,izhikevichbook,ashwin2016mathneuro,Parker2013}. Neuronal networks dynamically self-organize in various length and time-scales to give rise to a wide repertoire of behaviours. While the dynamical state of an individual neuron depends on the type of ion channels, their dendritic structure and the nature of coupling with other neurons, the complex behavioural aspects produced by the brain are due to the collective dynamics of several ensembles of neurons \cite{womelsdorf2007modulation,belykh2011mesoscale}. \par

In this context, community structured networks are one of the most commonly found network topology in the brain. Such networks consist of nodes that are grouped into communities based on some characteristics of the nodes or based on their connections with other nodes \cite{newman2006modularity,stroud2015dynamics}. There has been a lot of connectomics study which has pointed to the community structured or modular organization of the networks in the mammalian brain \cite{meunier2010modular,gleiser2010heir,nicolini2016modular}. Some notable examples of such architecture are the striatal networks, visual cortex and olfactory lobes in insects \cite{assisi2011structure} and also the layers of neurons in medial entorhinal cortex \cite{honey2007network,angulo2016cell}. Various studies have shown that this type of structure may have some role in the processing of information in neural circuits \cite{kumar2010spiking}. Some theoretical studies also explore the potential aspects of the modular organization leading to interesting dynamical phenomena in neuronal networks \cite{yang2017partial,hizanidis2016chimera,santos2017chimera}. \par

Several nonlinear dynamical models capture the dynamical features of individual neurons like spiking and bursts, the most widely used among them being Hindmarsh-Rose neuronal model \cite{hindmarsh1984model} that can demonstrate a broad class of bursting dynamics \cite{barrio2014macro,innocenti2007HR}. As is well established neurons can make contacts with other neurons through synapses that can be excitatory or inhibitory. Specifically, several studies report the effect of inhibitory synaptic connections in neurons \cite{van1994inhibition,lewis2003dynamics,assisi2012synaptic,borgers2003synchronization,catsigeras2010chaos}. This type of coupling is found very crucial in producing rhythmic activity in the brain as reported by various theoretical and experimental studies \cite{buzsaki2004,wang1996gamma,sohal2005inhibitory,white2000networks}. \par

Along with the nature of connectivity among the neurons, the major factor affecting the neuronal activity is the time-scale of its dynamics. This time-scale is generally dependent on the interplay of the various ions flowing across the cell membrane of the neuron deciding its capacitance and conductance \cite{echeveste2016drifting}. The collective activities of neurons are dependent on the temporal scales of activity of each neuron or ensemble of neurons \cite{rubin2008mmo,honey2007network,kispersky2010EHC,yamashita2008emergence,papo2013time} and they tend to play a major role especially in the oscillations in the midbrain region \cite{buzsaki2004}. \par

In this work, we address an important aspect of brain dynamics, viz how the presence of differing time-scales affect the emergent dynamics of the community structured neuronal networks. We start with the hypothesis that the presence of differing time-scales and inhibitory coupling can result in modulation of emergent frequencies in the system. We present a comprehensive model of a system where such variations in burst frequencies and emergence of frequency locked clusters can be studied in detail for a range of parameter values. This study in a way can model the modulation of oscillation frequencies and frequency locking reported from MEG/EEG data \cite{siebenhuhner2016cross} and intra-cortical recordings \cite{penn2016network} in primates and rats. Furthermore, such frequency locking can play a vital role in the formation of working memory.\par

The community structured network that we present is a very basic model that can address the dynamical complexity of the brain. It illustrates how a system of HR neurons can produce travelling burst sequences that circulate periodically among the modules. We observe a variety of dynamical states when the time-scales of the dynamics in the modules are varied. The most relevant among them are the frequency synchronized clusters with periodic sequences of bursting evident from the spatio-temporal dynamics of the network. \par

The model and methods of study are introduced in the next section. The results of our analysis on modular networks with identical neurons and those with time scale mismatch in their dynamics are detailed in the next three sections. The salient features and main results of the study are summarized in the final section.  \par


\section{Modular neuronal networks with differing time-scales}
\label{sec-2}
\begin{figure}
 \centering
 \begin{subfigure}[b]{0.4\textwidth}
   \includegraphics[width=\textwidth]{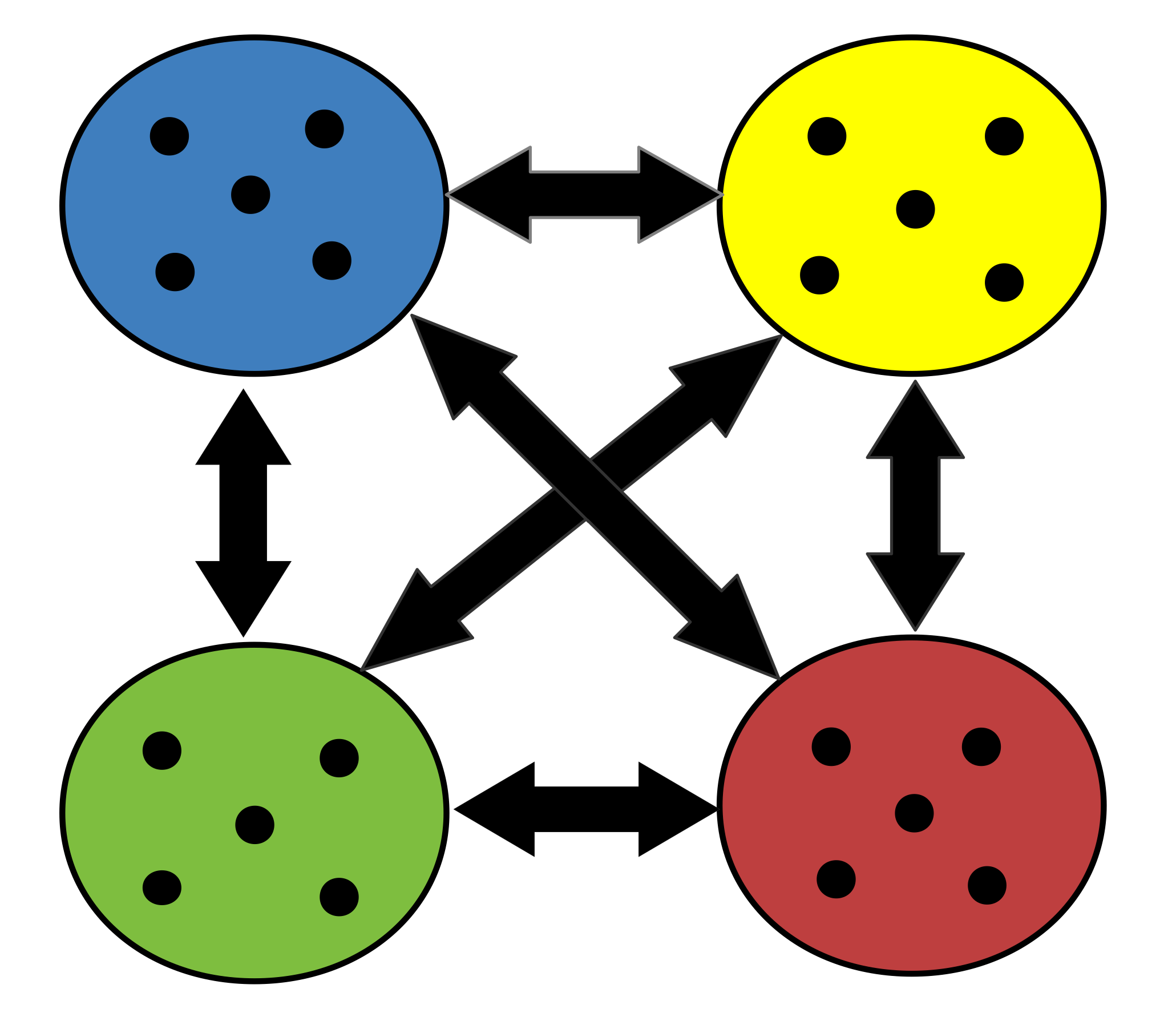}
   \caption{}
   \label{fig:f1a}
 \end{subfigure}
 \hfill
 \begin{subfigure}[b]{0.56\textwidth}
   \includegraphics[width=\textwidth]{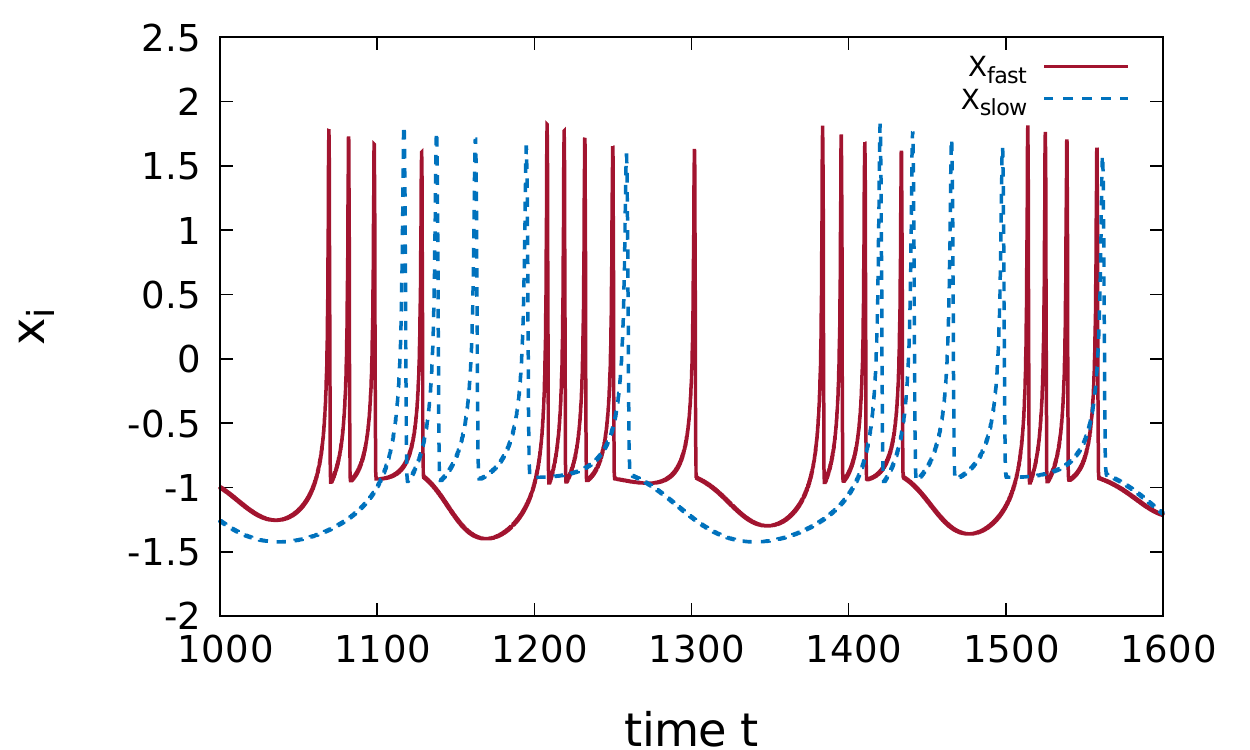}
   \caption{}
   \label{fig:f1b}
 \end{subfigure}
 \caption{(Colour Online) \textbf{(a)Schematic representation of a network with modular structure:} The circles represent the modules and the dots inside each circle represent a few typical neurons. The arrows represent the all-to-all coupling between neurons of two different modules. \textbf{(b)Intrinsic dynamics of HR neuron:} The chaotic bursts for parameters $I_e = 3$ and $\epsilon = 0.006$ for a neuron (smooth line) with ($\eta = 1$) and that for a slow neuron (dotted line) with time-scale mismatch $\eta = 0.5$}
 \label{fig:f1}
\end{figure}

We start with a model of commnity structured network consisting of $N$ neurons grouped into $M$ communities or modules. The neurons in each module belong to a community since they have a specific time-scale with zero intra-connections and full inter-connections. This means they do not have any connections among themselves but each of them has inhibitory connections with the other neurons of all the other communities. For this purpose, we construct the adjacency matrix $A$ with ($M \times M$) block matrices $B_{lk}$ of size $(N/M) \times (N/M)$ where $l,k = 1,2,..,M$. The required community structure is established by choosing the $B_{lk}$s as:
\begin{equation}
B_{lk} = 
\begin{cases}
[0],$ $ \forall $  $ l = k\\
[1],$ $ \forall $  $ l \neq k 
\end{cases}
\end{equation}
Thus the adjacency matrix, $A$, in this case is given as:
\begin{equation}
A = \begin{bmatrix}
    [0]&[1]&[1]&[1] \\
    [1]&[0]&[1]&[1] \\
    [1]&[1]&[0]&[1] \\
    [1]&[1]&[1]&[0] 
   \end{bmatrix}
\end{equation}
The elements of the adjacency matrix $A$ are such that, $a_{ij} = 0$ if the neurons are in the same block($ l = k$) and $a_{ij} = 1$ if they are in different blocks ($ l \neq k$). The schematic diagram in \autoref{fig:f1a} shows the structure thus chosen where the modules with the uncoupled neurons are represented by the colored circles. The typical values of the size of the network used in the study are $N=120$ and $M = 4$. \par

The intrinsic dynamics of each neuron is same as HR neuron and the dynamics of the $i^{th}$ neuron as:
\begin{multline}
\label{eq:whole}
\dot{x_i} = \eta_i \bigg(y_i - {x_i}^3 + 3{x_i}^2 - z_i + I_e - \beta(V - x_i)\sum_{j=1}^{N}a_{ij}\Big(\frac{1}{e^{-\lambda(x_j + K)}} \Big) \bigg) \\
\dot{y_i} = \eta_i(1 - 5{x_i}^2 - y_i) \\ 
\dot{z_i} = \eta_i(\epsilon( 4(x_i + x_r) - z_i)) \\
\end{multline}
Here, the variable $x_i$ represents the membrane potential and the variable $y_i$ and $z_i$ represent the fast and slow gating variables of the HR neuron. We use the parameters $I_e = 3$ and $\epsilon = 0.006$ such that the intrinsic dynamics shows chaotic bursts. The synaptic coupling function is a sigmoidal function with parameters $\lambda = 10$, reversal potential $V = 2$ and synaptic threshold $K = 0.25$ respectively. This type of coupling is chosen because it is more prevelant in the brain and generates nonintuitive emergent phenomena. The nature of coupling is inhibitory and the parameter $\beta$ gives the strength of coupling. \par

We introduce additionally a time-scale parameter determining the dynamics of each neuron as $\eta_i$, with values in the range $0 < \eta_i \leq 1$. It is clear that $\eta_i =1$ corresponds to the regular neuronal time-scales while any values of $\eta_i < 1$ would mean the corresponding neuron being slower in dynamics than others while keeping the nature of its intrinsic dynamics the same \cite{gupta2016suppression}. Biologically the parameter $\eta$ can be attributed to the differences in the capacitance of the axonal membranes. In \autoref{eq:whole}, the parameter $\eta$ can be used to tune the frequency of the neurons without affecting the dynamics of the neurons. \autoref{fig:f1b} shows the dynamics of two HR neurons, one of which has a slower time-scale with $\eta= 0.5$.  \par

In this study we keep the time scale of all the neurons in each module at a specific value of the time-scale $\eta_i$. So for the whole network, time-scale parameters form a set of $M$ values, $\boldsymbol{\eta}= (\eta_1, \eta_2, \eta_3, \eta_4)$. We initially consider a network that is symmetric both in structure and dynamics with identical neurons with all modules having the same time scale i.e. $\boldsymbol{\eta}_0 = \irow{1, 1, 1, 1}$. In this case we analyze the emergent burst patterns of the whole network using spatio-temporal plots.  Then we proceed to the more interesting case where one level of symmetry is broken by having two modules follow a slower time scale with a mismatch with the faster ones given by $\boldsymbol{\eta}_1 = \irow{1, \eta, 1, \eta}$ (where $\eta \in [0,1)$). We also study two other possible configurations with $\boldsymbol{\eta}_2 = \irow{\eta, \eta, \eta, 1}$ and $\boldsymbol{\eta}_3 = \irow{1, 1, 1, \eta}$.\par

In all the cases, the system of equations in \autoref{eq:whole} is analyzed by numerical simulations using fourth order vector Runge-Kutta algorithm for a step $\Delta t = 0.01$ and $600000$ iterations. The first $100000$ values are discarded as transients and the remaining are used in the analysis. We present the spatio-temporal plots that provide a qualitative picture of the emergent dynamics of the network. \par

For a detailed characterization of the temporal dynamics, we compute the burst frequency of each neuron from its time series, $x_i$. We note the time ${\tau_i}^k$ at which the $x_i$ values cross a threshold value of $-1$ with the $\dot{x} > 0$. The value of the threshold is carefully chosen after studying several cases with different parameter values for each neuron. The ${\tau_i}^k$ represents the time of onset of the $k^{th}$ burst in the $i^{th}$ neuron. The inter-burst interval (IBI) is then calculated for each burst as the time interval $\Delta \tau^k = \tau^{k+1} - \tau^{k}$. The average burst frequency of the $i^{th}$ neuron is calculated using the equation: 
\begin{equation}
\label{eqn:3}
f_i = \dfrac{2\pi}{K_i} \sum_{k=1}^{K_i} \dfrac{1}{ {\tau_i}^{k+1} - {\tau_i}^k } 
\end{equation}      
where $K_i$ refers to the total number of bursts for the $i^{th}$ neuron in the total time used for calculation. The intrinsic burst frequency of a single HR neuron calculated using this method turns out to be $ f_{o} \backsimeq 0.04$ in arbitrary units. This generally corresponds to a frequency of $\backsim 60$ Hz when we compare it with the biological neuronal time series \cite{hindmarsh1984model}. \par

\section{Travelling burst dynamics}
\label{sec-3}
\begin{figure}[h]
\centering
\begin{subfigure}[h]{0.48\textwidth}
   \includegraphics[width=\textwidth]{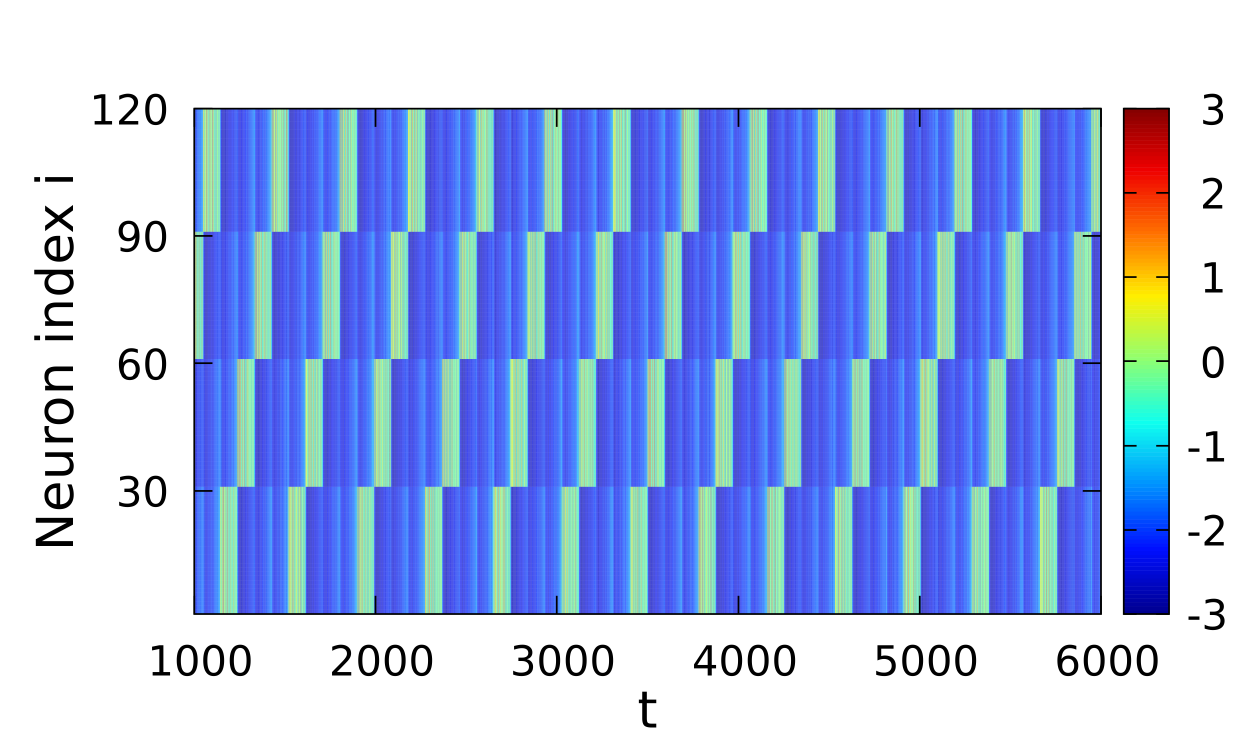}
   \caption{$\beta = 0.1$}
   \label{fig:f2a} 
\end{subfigure}
\hfill
\begin{subfigure}[h]{0.48\textwidth}
   \includegraphics[width=\textwidth]{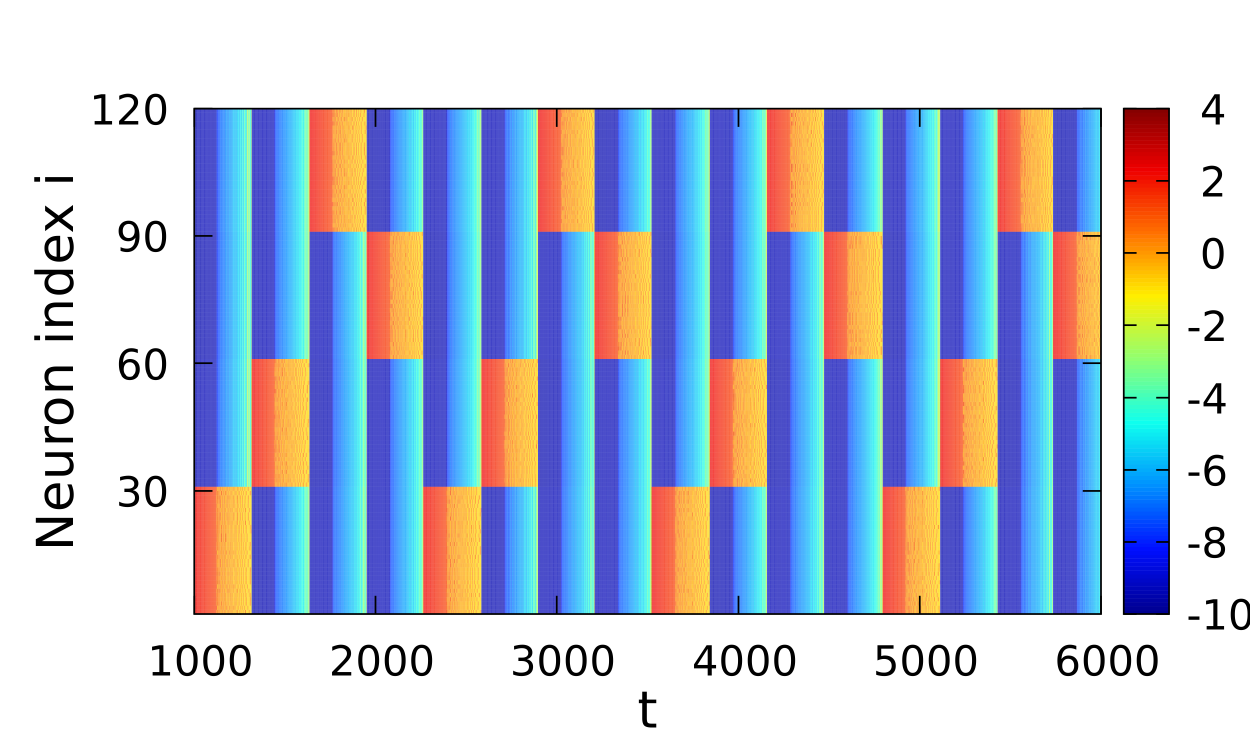}
   \caption{$\beta = 2$}
   \label{fig:f2b}
\end{subfigure}
\caption{(Colour Online) \textbf{Spatio-temporal plots of the neuronal dynamics:}(a) $\beta = 0.1$ , (b) $\beta = 2$. Each module of neurons bursts sequentially one after the other in time, producing a travelling burst pattern in the network. }
\label{fig:f2}
\end{figure}

The type of topology and coupling considered in the study leads to synchronization between uncoupled neurons due to the action of inhibitory inputs from other neurons. Thus all neurons in a particular community or module are in complete synchrony with each other. In the case of two neurons, it is known that inhibitory nature of the coupling leads to anti-phase synchronization because, when one of the neurons is active it tends to inhibit the other neuron's activity. The same tendency in a modular network of mutually coupled inhibitory neurons can drive the neurons to a state of sequential bursting. So when we look at the bursting pattern in the whole network, we see each module of neurons bursts one after the other in a specific sequence like travelling bursts. The spatio-temporal plots in \autoref{fig:f2} illustrate these emergent bursting patterns in the network for two different coupling strengths $\beta$.  \par

\begin{figure}[h]
\centering
\begin{subfigure}[h]{0.48\textwidth}
   \includegraphics[width=\textwidth]{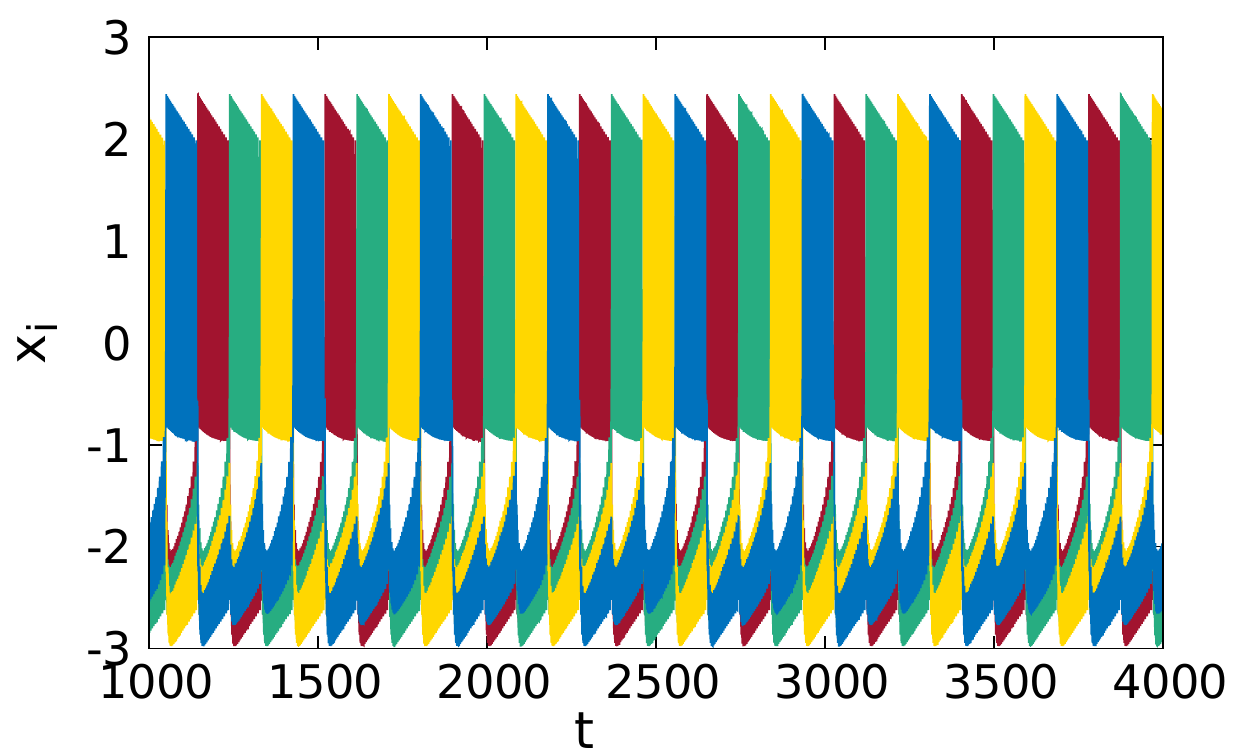}
   \caption{$\beta = 0.1$}
   \label{fig:f3a} 
\end{subfigure}
\hfill
\begin{subfigure}[h]{0.48\textwidth}
   \includegraphics[width=\textwidth]{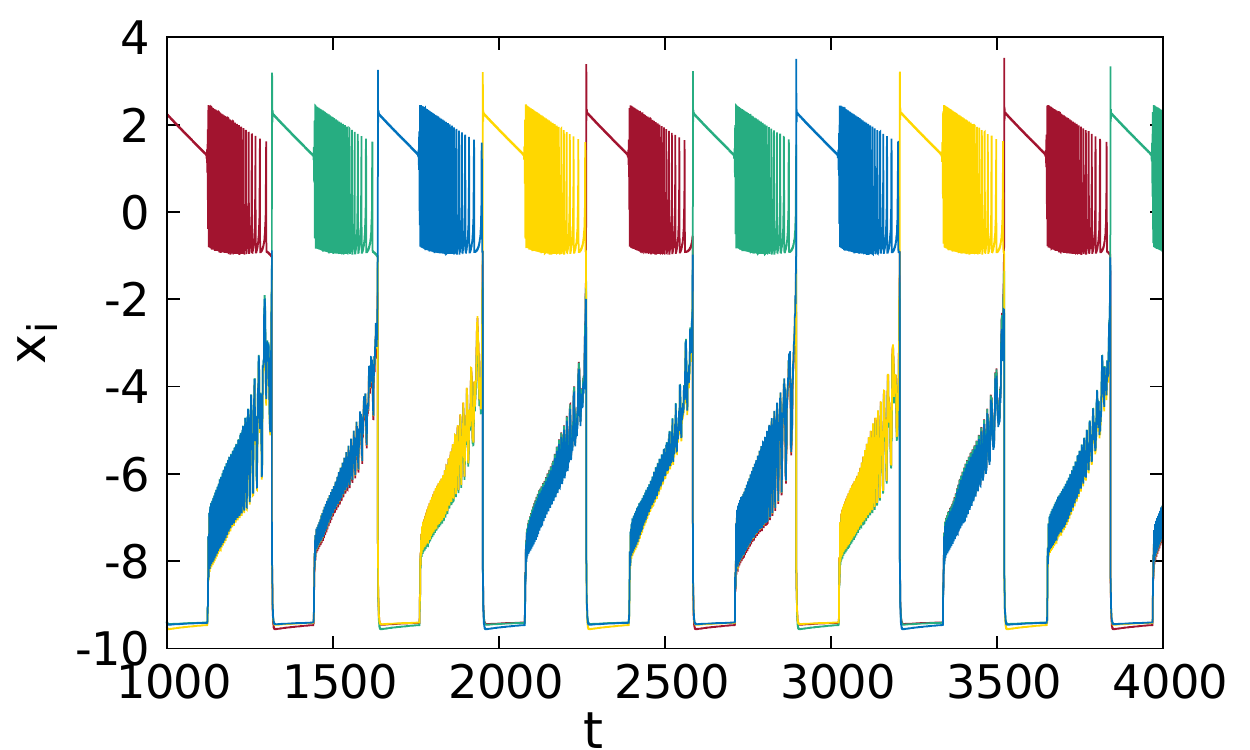}
   \caption{$\beta = 2$}
   \label{fig:f3b}
\end{subfigure}
\caption{(Colour Online) \textbf{Time series of sequentially bursting neurons from each module:}(a) $\beta = 0.1$ , (b) $\beta = 2$. The dynamics of typical neurons from each module ( for $i = 30, 60, 90, 120$) is shown. It is clear that all neurons have the same periodic bursts that form a sequential pattern of bursting. With increase in $\beta$, the time period of the burst increases. }
\label{fig:f3}
\end{figure}

We can infer a few general trends in the dynamics of the network from the spatio-temporal plots in \autoref{fig:f2}. Firstly, the bursting sequence remains periodic as the whole system evolves in time. This means the inhibitory coupling drives the system of chaotic bursting neurons into periodic square bursting dynamics. Secondly, the bursting dynamics is identical in all the modules, i.e., all the neurons show identical mixed-mode square bursting dynamics, although the bursts are shifted in space and time. This is illustrated in \autoref{fig:f3} where the time series of typical neurons from each module corresponding to the states of spatio-temporal plots in \autoref{fig:f2} are shown.\par

\begin{figure}[!htb]
\centering
\includegraphics[width=0.6\textwidth]{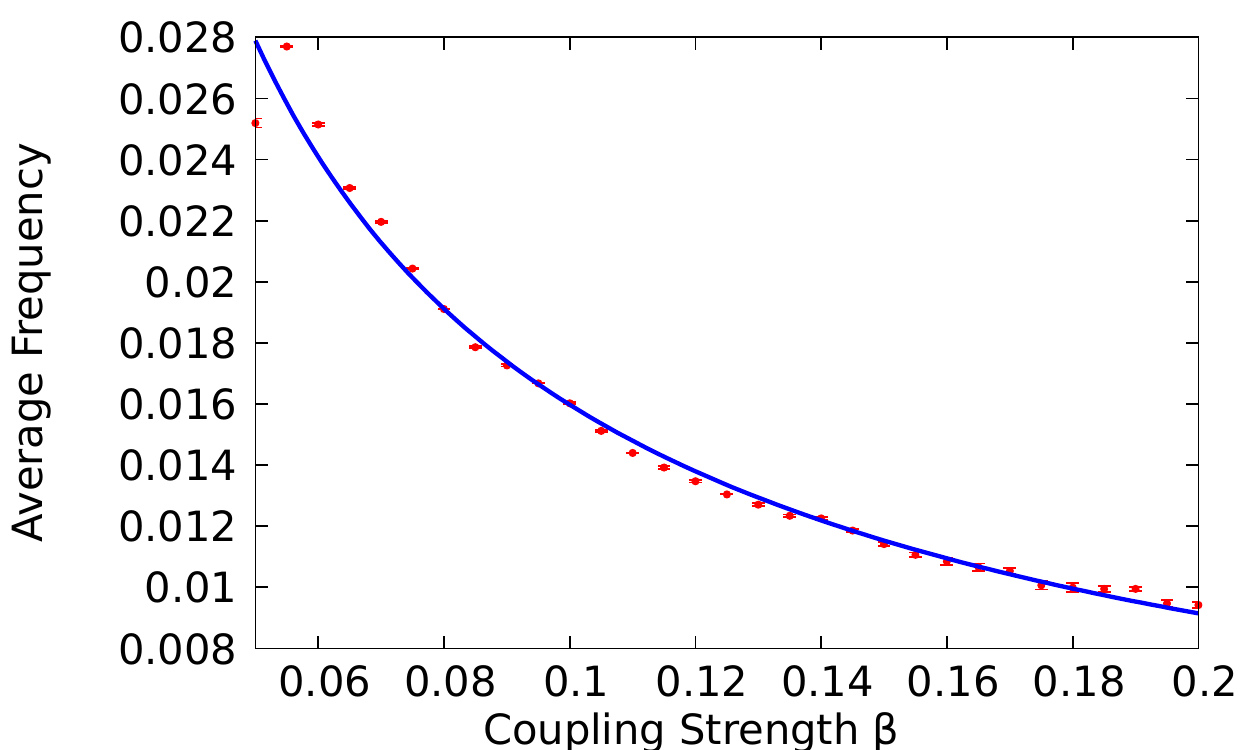}
\caption[shrink=12]{(Colour Online) \textbf{Frequency vs Coupling Strength:} The dotted circles indicate the calculated frequency averaged over the network for five realizations. The smooth line shows the function fit to the mean values, $ f_{av} = A\beta^{-\nu}$ where the values are given as $A = 0.0025 \pm 0.0001$ and $\nu = 0.80 \pm 0.02$.}
\label{fig:f4}
\end{figure}

We calculate the burst frequency of each neuron averaged over the module, using \autoref{eqn:3}. This remains the same for all the modules for a given $\beta$, indicating frequency synchronization over the whole network, even though the bursts are shifted in phase. Moreover, as is clear from the \autoref{fig:f3}, the bursting frequency decreases as $\beta$ increases. In the context of neurons, the synaptic coupling strength $\beta$ can vary according to certain plasticity rules \cite{abbott2000synaptic}. Our results indicate how this variation in $\beta$ can modulate the frequency of the collective dynamics of neurons. This is shown in \autoref{fig:f4} where $f_{av} \simeq \beta^{-\nu}$. We note that frequency of the whole network is less than the intrinsic frequency of the neurons. For the case of the symmetric network discussed so far, with identical neurons, with $\boldsymbol{\eta}_0 = \irow{1, 1, 1, 1}$, all the neurons are frequency synchronized and the firing pattern has a wave of travelling bursts circulating over the whole network periodically.  \par

\section{Bursts sequences with differing time-scales}
\label{sec-4}
In this section, we present the study on the modular network with a time-scale parameter is not the same in all the modules. The differing time-scales break the symmetry of the network, as now the neurons in one module are non-identical with that in another. As a specific case we consider, two slow communities and two fast ones with the time-scales fixed as $\boldsymbol{\eta}_1 = \irow{1, \eta, 1, \eta}$. \par

\begin{figure}[h]
\centering
\begin{subfigure}{0.48\textwidth}
\includegraphics[width=\textwidth]{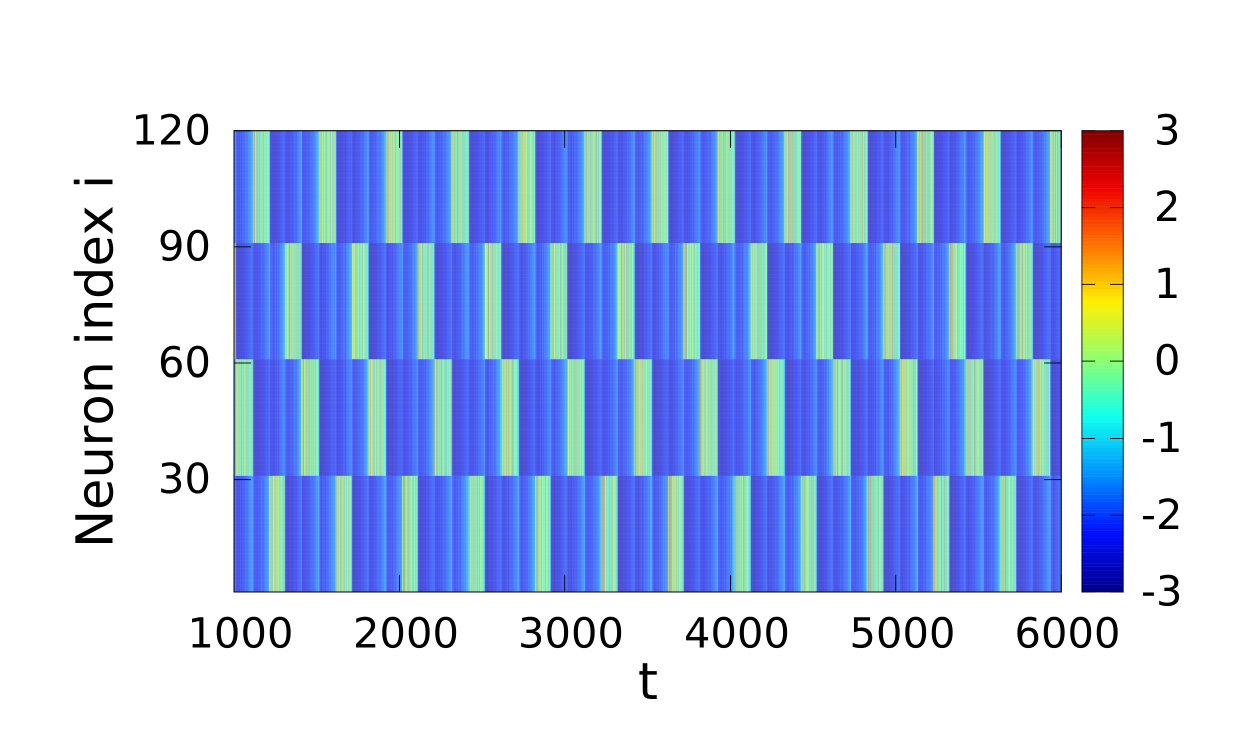}
\caption{$\eta = 0.9$}
\label{fig:f5a}
\end{subfigure}
\hfill
\begin{subfigure}{0.48\textwidth}
\includegraphics[width=\textwidth]{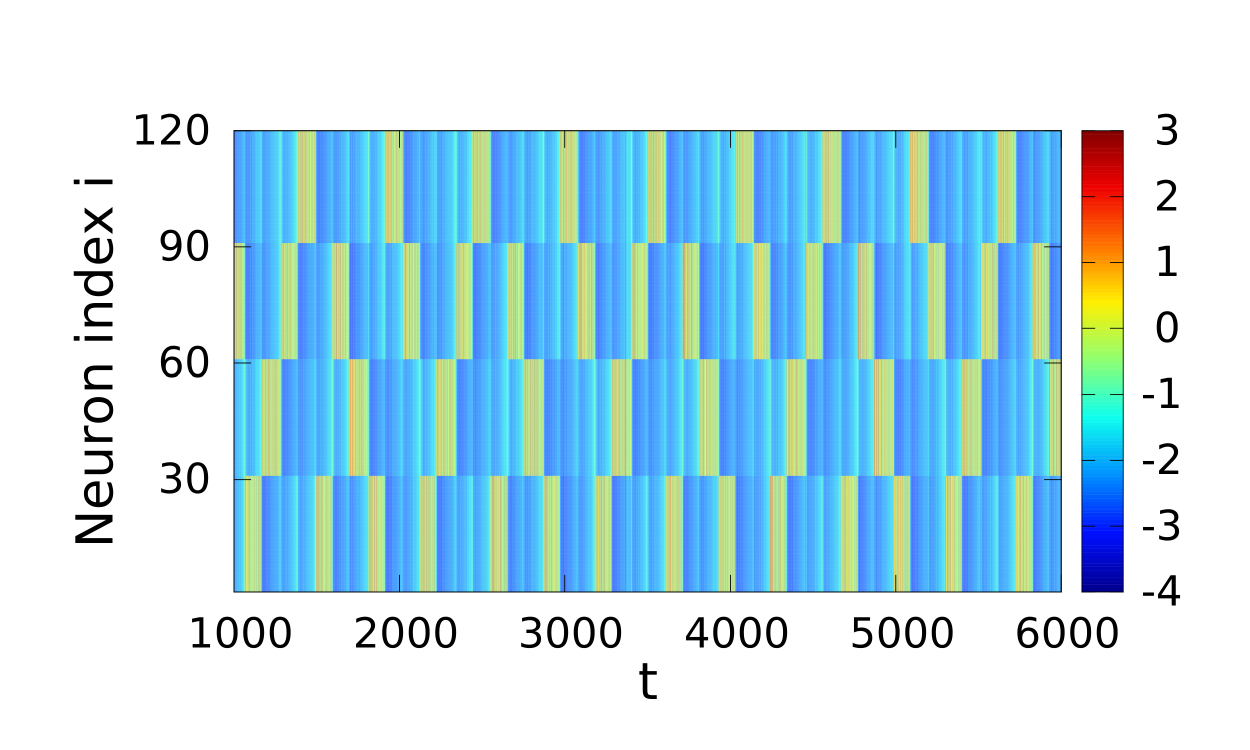}
\caption{$\eta = 0.8$}
\label{fig:f5b}
\end{subfigure}
\hfill
\begin{subfigure}{0.48\textwidth}
\includegraphics[width=\textwidth]{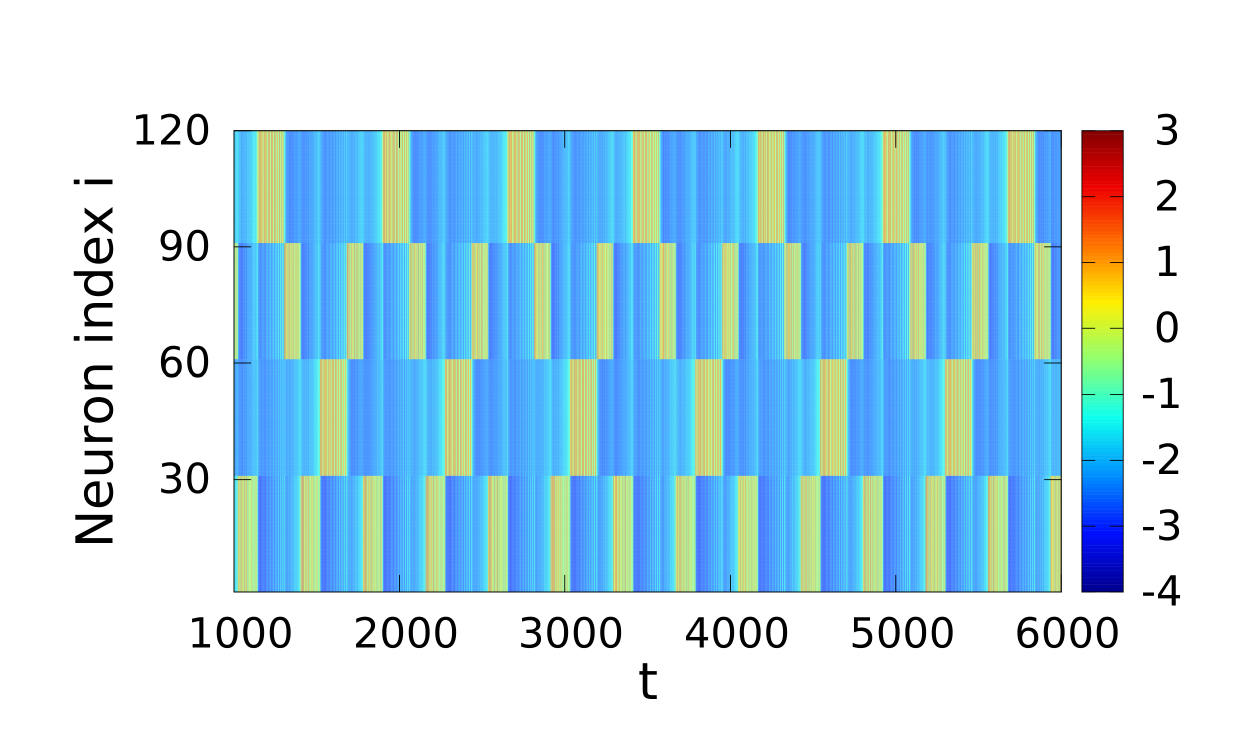}
\caption{$\eta = 0.4$}
\label{fig:f5c}
\end{subfigure}
\caption{(Colour Online) \textbf{Spatio-temporal plots of the neuronal dynamics with time-scale mismatch at $\beta = 0.1$:} (a)$\eta = 0.9$, showing the sequence $P^{\boldsymbol{\eta}_1}_1$. (b)$\eta = 0.8$, $P^{\boldsymbol{\eta}_1}_2$. (c)$\eta = 0.4$, $P^{\boldsymbol{\eta}_1}_3$.}
\label{fig:f5}
\end{figure}

Our numerical analysis shows that in this case also, the travelling burst pattern occurs with the neurons in each module bursting sequentially one after the other. But the pattern of bursting is decided by the value of the time-scale $\eta$. This is evident from the spatio-temporal dynamics on the network as shown in \autoref{fig:f5}, for three different values of $\eta$. We label each particular sequence by $P^{\boldsymbol{\eta}}_k$. For example, for $\eta = 0.9$, we identify the sequence of the travelling bursts from \autoref{fig:f5a} as $P^{\boldsymbol{\eta}_1}_1 = \overline{S_1S_2F_1F_2}$, where $S_1,S_2$ and $F_1,F_2$ represent the slow and fast community, numbered according to the order in which communities burst in time. The bar over the sequence indicates that the sequence is repeated periodically in time. \par

As $\eta$ is lowered to a value of $\eta = 0.8$, the travelling bursts have a longer sequence  $P^{\boldsymbol{\eta}_1}_2 =\overline{F_1S_1F_2S_2F_1F_2S_1F_1S_2F_2}$ with slow communities bursting four times and fast communities bursting six times during the time of one cycle, as shown in \autoref{fig:f5b}. At lower values of $\eta$, like in \autoref{fig:f5c}, ($\eta = 0.4$) the bursting pattern changes to $P^{\boldsymbol{\eta}_1}_3 = \overline{S_2F_2F_1S_1F_2F_1}$ so that the slower modules burst once each and the faster modules burst twice in one cycle. In all cases, the sequences are repeated periodically in time. We prescribe a characterization of the periodic bursting sequences as $(p:q)$ where $p$ represent the number of slow bursts and $q$ represent the number of fast bursts. Then, with time scales fixed as $\boldsymbol{\eta}_1 = \irow{1, \eta, 1, \eta}$, the network has $3$ different emergent dynamical states of the network, that can be denoted by $(2:2)$, $(4:6)$ and $(2:4)$ for decreasing values of $\eta$. \par

\begin{figure}[!htb]
\centering
\includegraphics[width=0.6\textwidth]{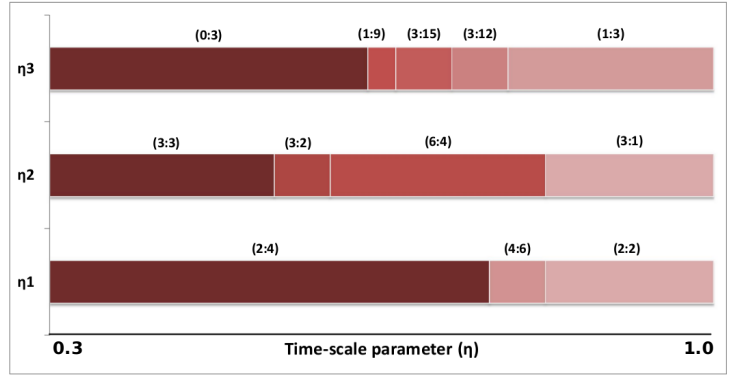}
\caption[shrink=12]{(Colour Online) \textbf{Schematic representation of the burst sequences observed in the three configurations of the modules for $\beta = 0.1$ : }For each configuration, the different sequences are in different colours with $(p:q)$ values given above.}
\label{fig:f6}
\end{figure}

We now present two more possible configurations of the network that are more heterogeneous with unequal number of slow and fast modules, with 3 slow and 1 fast modules ($\boldsymbol{\eta}_2 = \irow{\eta, \eta,\eta, 1}$), and with 1 slow and 3 fast modules ($\boldsymbol{\eta}_3 = \irow{1, 1, 1, \eta}$). In a network of four modules, these exhaust all possible configurations. We have summarized these results in \autoref{table:1} and \autoref{fig:f6} highlighting the ranges of $\eta$ values for each burst sequence for all the three cases in the present study. By extending the study along similar lines to networks with larger number of modules, we expect a variety of interesting periodic patterns of bursting based on the numbers of slow and fast modules and the distribution of $\eta$ values among them. \par

\begin{table}
\begin{tabular}{|p{3cm}|p{7cm}|p{1.5cm}|p{2cm}|  } 
  \hline 
 \multicolumn{4}{|c|}{Burst sequences in four modules}\\
 \hline 
 Configuration ($\boldsymbol{\eta}_i$)& Sequnce ($P^{\boldsymbol{\eta}_i}_j$) & $(p:q)$ & Range of $\eta$\\
 \hline 
 \multirow{4}{*}{$\boldsymbol{\eta}_1 = \irow{1, \eta, 1, \eta}$} & $P^{\boldsymbol{\eta}_2}_1 = \overline{S_3S_2S_1F}$ & $(3:1)$ & $[0.83,1]$\\  
 & $P^{\boldsymbol{\eta}_1}_2 =\overline{F_1S_1F_2S_2F_1F_2S_1F_1S_2F_2}$ & $(4:6)$ & $(0.76,0.82]$\\ 
 & $P^{\boldsymbol{\eta}_1}_3 = \overline{S_2F_2F_1S_1F_2F_1}$ & $(2:4)$ & $[0.4,0.76]$\\ 
\hline 
 \multirow{5}{*}{$\boldsymbol{\eta}_2 = \irow{\eta, \eta, \eta, 1}$} & $P^{\boldsymbol{\eta}_2}_1 = \overline{FS_3FS_2FS_1}$ & $(3:3)$ & $[0.83,1]$\\  
 & $P^{\boldsymbol{\eta}_2}_2 = \overline{FS_3S_2FS_1S_3FS_2S_1F}$ & $(6:4)$ & $[0.6,0.83)$\\ 
 & $P^{\boldsymbol{\eta}_2}_3 = \overline{S_3FS_2S_1F}$ & $(3:2)$ & $[0.54,0.6)$\\
 & $P^{\boldsymbol{\eta}_2}_4 = \overline{FS_3FS_2FS_1}$ & $(3:3)$ & $[0.3,0.54)$\\ 
 \hline
 \multirow{6}{*}{$\boldsymbol{\eta}_3 = \irow{1, 1, 1, \eta}$} & $P^{\boldsymbol{\eta}_3}_1 = \overline{F_3F_2F_1S}$ & $(1:3)$ & $(0.78,1]$\\  
 & $P^{\boldsymbol{\eta}_3}_2 = \overline{SF_3F_1F_2F_3SF_1F_2F_3F_1SF_2F_3F_1F_2}$ & $(3:12)$ & $(0.72,0.78]$\\ 
 & $P^{\boldsymbol{\eta}_3}_3 = \overline{SF_1F_3F_2F_1F_3SF_2F_1F_3F_3F_1SF_3F_2F_1F_3F_2}$ & $(3:15)$ & $(0.66,0.72]$\\ 
 & $P^{\boldsymbol{\eta}_3}_4 = \overline{SF_3F_2F_1F_3F_2F_1F_3F_2F_1}$ & $(1:9)$ & $(0.63,0.66]$\\
 & $P^{\boldsymbol{\eta}_3}_5 = \overline{F_3F_2F_1}$ & $(0:3)$ & $[0.3,0.63]$ \\
 \hline
\end{tabular}
\caption{Burst patterns and their sequences for the three configurations with four modules for $\beta = 0.1$. The range of time-scale values for which they are stable are also given in the rightmost column. }
\label{table:1}
\end{table}

\section{Frequency locked clusters}
\label{sec-5}
\begin{figure}[h]
\centering
\begin{subfigure}[h]{0.44\textwidth}
   \includegraphics[width=\textwidth]{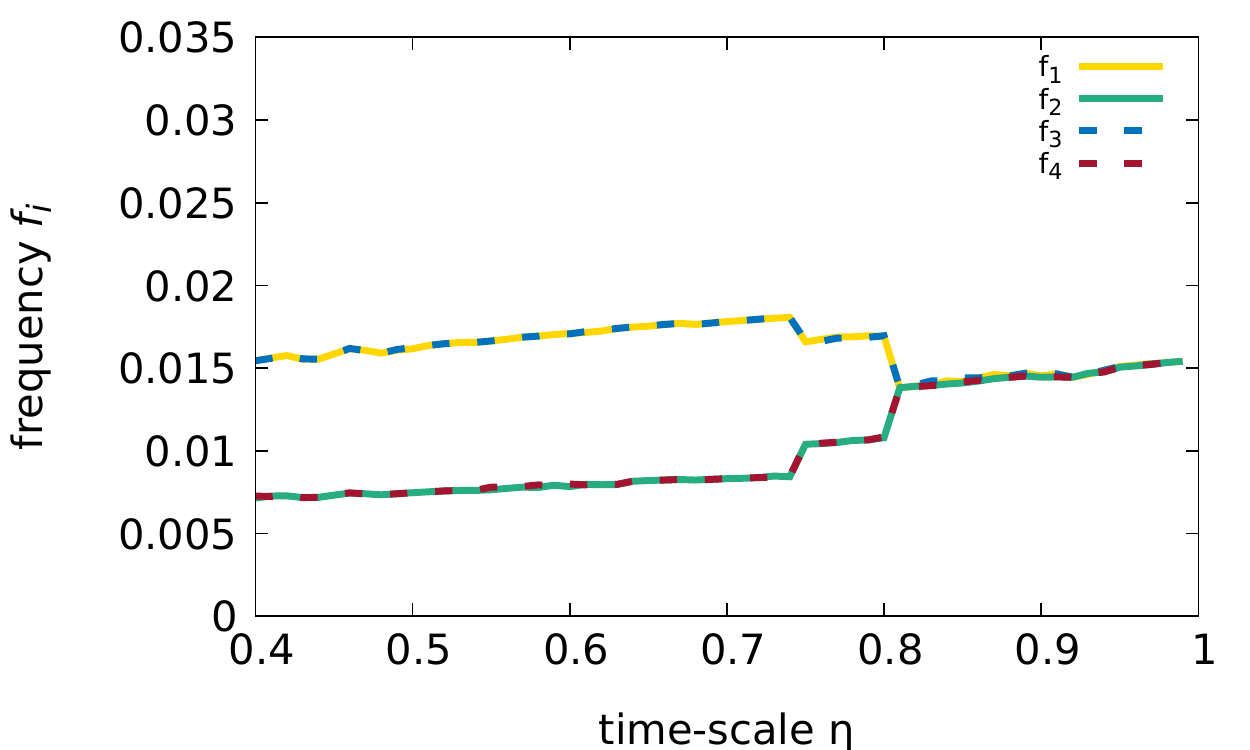}
   \caption{ }
   \label{fig:f7a}
\end{subfigure}
\hfill
\begin{subfigure}[h]{0.5\textwidth}
   \includegraphics[width=\textwidth]{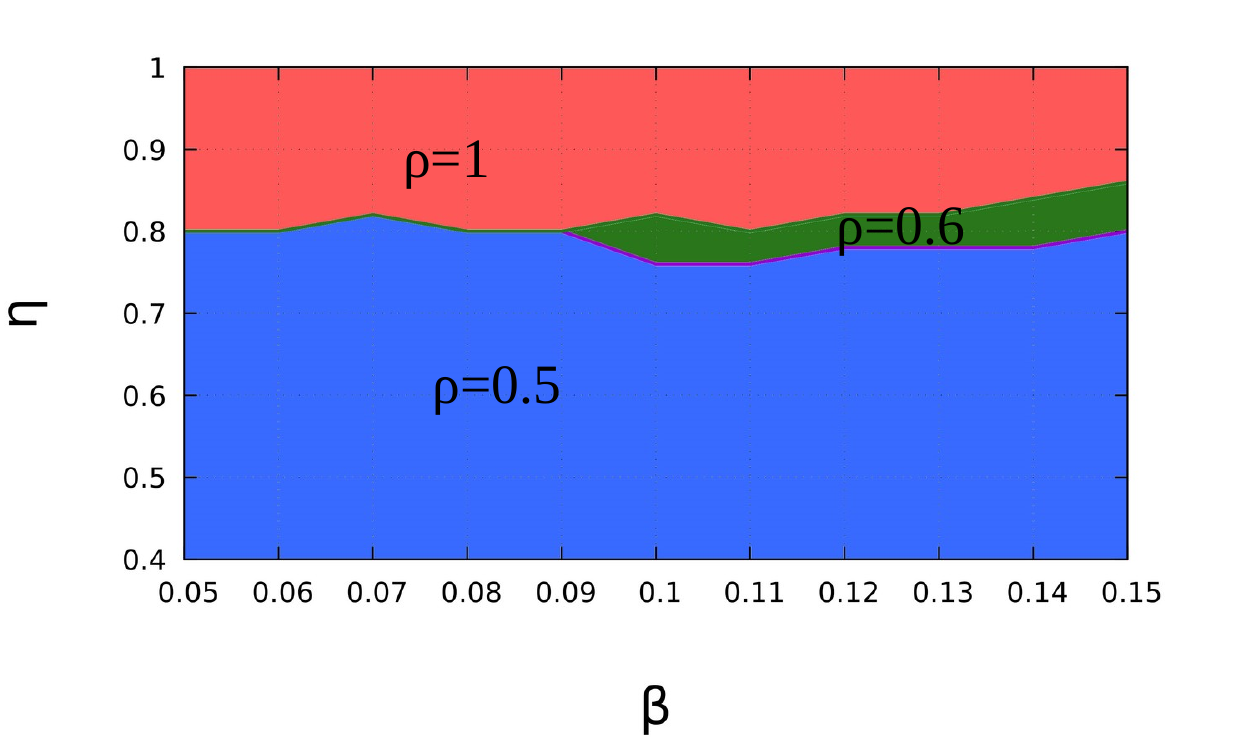}
   \caption{ }      
   \label{fig:f7b}
\end{subfigure}
\caption{(Colour Online) \textbf{(a)Frequency vs time-scale parameter at $\beta = 0.1$ for configuration $\boldsymbol{\eta}_1$:} For large $\eta$ or less mismatch in time-scale the whole network is synchronized at a single frequency. At lower values of $\eta$, there are two distinct clusters of slow and fast frequencies. \textbf{(b)Regions in the ($\beta,\eta$) parameter space for configuration $\boldsymbol{\eta}_1$:}The three regions correspond to the three distinct synchronized frequency locked states - Red(gray) region corresponds to the state $\rho = 1$, green(black) region to $\rho = 0.6$ and blue(dark gray) region to $ \rho = 0.5$.}
\label{fig:f7}
\end{figure}

The pattern of sequences reported in \autoref{sec-4} can be further quantified by calculating the average burst frequency for each community of neurons as $\boldsymbol{f} = \irow{f_1, f_2, f_3, f_4}$ using \autoref{eqn:3}. For the case of two slow and two fast modules, the frequencies for different $\eta$ values with $\beta=0.1$ for each module is shown in \autoref{fig:f7a}. It is clear that for sufficiently small mismatch or large $\eta$, all the modules in the network are still in a single frequency synchronized state with the average frequency much less than the intrinsic value. As $\eta$ is reduced, we find the single frequency state bifurcates into a state where the two slow and the two fast modules separate into two different clusters of differing frequencies, while being synchronized among them. \par

We find that the bursts in the slow and fast modules are locked into specific ratios. This ratio $\rho$, of the average frequency of neurons in the slow modules and that of neurons in the fast modules, can thus characterize the locked states. For the network with two slow and two fast modules, we get three distinct ratios, $1$, $0.5$ and $0.6$. The values of the parameters for which the locked states stabilize, are indicated in the parameter plane $(\beta$, $\eta)$ in \autoref{fig:f7b}. It is interesting to note that while the frequency values of each cluster keep changing with $\beta$ and $\eta$, the locking ratio remains constant over a sufficient region of the plane. \par
  
These frequency locked states correspond directly to the three patterns of spatial burst sequences mentioned above and seen in the spatio-temporal plots in \autoref{fig:f5}. Thus $\rho = 1$ corresponds to the state $(2:2)$, $\rho = \frac{1}{2}$ corresponds to state $(2:4)$, and $\rho = \frac{2}{3}$ corresponds to the state $(4:6)$. Thus the frequency ratio and the burst sequences together characterize the spatio-temporal dynamics of the neurons fully. \par

\section{Discussions}
\label{sec-6}
In this study we report the emergent dynamics and bursting patterns produced in a modular network of neurons with inhibitory couplings. The modules in the network are organized based on the type of coupling and the dynamical time-scales. Primarily our study emphasizes the role of inhibitory couplings in inducing regulated patterns of synchrony among neurons which are not coupled directly to each other. It also indicates how the frequency of the whole network gets modified due to variations in the synaptic coupling strength. \par

We show that a variety of dynamical states are possible when the time-scales of the dynamics in the communities are different. The travelling burst dynamics observed in this case is specific to the mismatch in time-scales and can be uniquely characterized by the frequency ratios. The slow and fast time-scales in each module result in specific emergent frequencies in it, which in the presence of inhibitory couplings among them, drive the whole network to fire at various patterns of bursting sequences. The resultant spatio-temporal regularity in the collective dynamics is fully characterized by analyzing the sequential order in the burst patterns and the frequency locking in temporal order. By changing the time-scale mismatch parameter, we observe frequency synchronized clusters that are locked into ratios like $1$, $\frac{1}{2}$ and $\frac{2}{3}$, etc. and are stable over a finite region of the parameter plane of time-scale and coupling strength. \par

As is well known, the frequency of bursting or firing in the dynamics of connected neurons has a major role in the coding and transmission of information among them \cite{borst1999information,averbeck2004coding,stein1967information} especially in selective communication \cite{izhikevich2003bursts}. Also, frequency locking and modulation are associated with integrating sensory information to form working memory and retrieval of such memory for sensory-guided tasks \cite{tiesinga2009cortical,womelsdorf2007attention}. In response to input currents, synaptic strengths, and heterogeneous synapses, modulation in frequency and frequency locked states have been reported earlier\cite{stoop2004phase,lowet2015input,chow1998frequency}. But frequency locking and bursting sequences combining temporal and spatial dynamics due to heterogeneous time-scales is a novel phenomenon reported in this work. \par

By extending the model to a larger number of modules and varying the time scale mismatches among them, a variety of sequences can be generated. These bursting sequences themselves can thus be a mechanism or medium for coding information.\par

\section{References}
\label{sec-7}

\begin{thebibliography}{10}
\expandafter\ifx\csname url\endcsname\relax
  \def\url#1{\texttt{#1}}\fi
\expandafter\ifx\csname urlprefix\endcsname\relax\def\urlprefix{URL }\fi
\expandafter\ifx\csname href\endcsname\relax
  \def\href#1#2{#2} \def\path#1{#1}\fi

\bibitem{bullmore2009complex}
E.~Bullmore, O.~Sporns, Complex brain networks: graph theoretical analysis of
  structural and functional systems., Nature Reviews Neuroscience 10~(3).

\bibitem{izhikevichbook}
E.~M. Izhikevich, Dynamical systems in neuroscience, MIT press, 2007.

\bibitem{ashwin2016mathneuro}
P.~Ashwin, S.~Coombes, R.~Nicks, Mathematical frameworks for oscillatory
  network dynamics in neuroscience, The Journal of Mathematical Neuroscience
  6~(1) (2016) 2.

\bibitem{Parker2013}
D.~Parker, V.~Srivastava, Dynamic systems approaches and levels of analysis in
  the nervous system, Frontiers in Physiology 4.

\bibitem{womelsdorf2007modulation}
T.~Womelsdorf, J.-M. Schoffelen, R.~Oostenveld, W.~Singer, R.~Desimone, A.~K.
  Engel, P.~Fries, Modulation of neuronal interactions through neuronal
  synchronization, Science 316~(5831) (2007) 1609--1612.

\bibitem{belykh2011mesoscale}
I.~Belykh, M.~Hasler, Mesoscale and clusters of synchrony in networks of
  bursting neurons, Chaos: An Interdisciplinary Journal of Nonlinear Science
  21~(1) (2011) 016106.

\bibitem{newman2006modularity}
M.~E. Newman, Modularity and community structure in networks, Proceedings of
  the National Academy of Sciences 103~(23) (2006) 8577--8582.

\bibitem{stroud2015dynamics}
J.~Stroud, M.~Barahona, T.~Pereira, Dynamics of cluster synchronisation in
  modular networks: Implications for structural and functional networks, in:
  Applications of Chaos and Nonlinear Dynamics in Science and Engineering-Vol.
  4, Springer, 2015, pp. 107--130.

\bibitem{meunier2010modular}
D.~Meunier, R.~Lambiotte, E.~T. Bullmore, Modular and hierarchically modular
  organization of brain networks, Frontiers in Neuroscience 4.

\bibitem{gleiser2010heir}
P.~M. Gleiser, V.~I. Spoormaker, Modelling hierarchical structure in functional
  brain networks, Philosophical Transactions of the Royal Society of London A:
  Mathematical, Physical and Engineering Sciences 368~(1933) (2010) 5633--5644.

\bibitem{nicolini2016modular}
C.~Nicolini, A.~Bifone, Modular structure of brain functional networks:
  breaking the resolution limit by surprise, Scientific Reports 6.

\bibitem{assisi2011structure}
C.~Assisi, M.~Stopfer, M.~Bazhenov, Using the structure of inhibitory networks
  to unravel mechanisms of spatiotemporal patterning, Neuron 69~(2) (2011)
  373--386.

\bibitem{honey2007network}
C.~J. Honey, R.~K{\"o}tter, M.~Breakspear, O.~Sporns, Network structure of
  cerebral cortex shapes functional connectivity on multiple time scales,
  Proceedings of the National Academy of Sciences 104~(24) (2007) 10240--10245.

\bibitem{angulo2016cell}
D.~Angulo-Garcia, J.~D. Berke, A.~Torcini, Cell assembly dynamics of
  sparsely-connected inhibitory networks: a simple model for the collective
  activity of striatal projection neurons, PLoS Computational Biology 12~(2).

\bibitem{kumar2010spiking}
A.~Kumar, S.~Rotter, A.~Aertsen, Spiking activity propagation in neuronal
  networks: reconciling different perspectives on neural coding, Nature
  reviews. Neuroscience 11~(9) (2010) 615.

\bibitem{yang2017partial}
X.~Yang, H.~Li, Z.~Sun, Partial coupling delay induced multiple spatiotemporal
  orders in a modular neuronal network, PloS one 12~(6) (2017) e0177918.

\bibitem{hizanidis2016chimera}
J.~Hizanidis, N.~E. Kouvaris, Z.-L. Gorka, A.~D{\'\i}az-Guilera, C.~G.
  Antonopoulos, Chimera-like states in modular neural networks, Scientific
  Reports 6.

\bibitem{santos2017chimera}
M.~Santos, J.~Szezech, F.~Borges, K.~Iarosz, I.~Caldas, A.~Batista, R.~Viana,
  J.~Kurths, Chimera-like states in a neuronal network model of the cat brain,
  Chaos, Solitons \& Fractals 101 (2017) 86--91.

\bibitem{hindmarsh1984model}
J.~L. Hindmarsh, R.~Rose, A model of neuronal bursting using three coupled
  first order differential equations, Proceedings of the Royal Society of
  London B: Biological Sciences 221~(1222) (1984) 87--102.

\bibitem{barrio2014macro}
R.~Barrio, M.~Angeles~Mart{\'\i}nez, S.~Serrano, A.~Shilnikov, Macro-and
  micro-chaotic structures in the hindmarsh-rose model of bursting neurons,
  Chaos: An Interdisciplinary Journal of Nonlinear Science 24~(2) (2014)
  023128.

\bibitem{innocenti2007HR}
G.~Innocenti, A.~Morelli, R.~Genesio, A.~Torcini, Dynamical phases of the
  hindmarsh-rose neuronal model: Studies of the transition from bursting to
  spiking chaos, Chaos: An Interdisciplinary Journal of Nonlinear Science
  17~(4) (2007) 043128.

\bibitem{van1994inhibition}
C.~Van~Vreeswijk, L.~Abbott, G.~B. Ermentrout, When inhibition not excitation
  synchronizes neural firing, Journal of Computational Neuroscience 1~(4)
  (1994) 313--321.

\bibitem{lewis2003dynamics}
T.~J. Lewis, J.~Rinzel, Dynamics of spiking neurons connected by both
  inhibitory and electrical coupling, Journal of Computational Neuroscience
  14~(3) (2003) 283--309.

\bibitem{assisi2012synaptic}
C.~Assisi, M.~Bazhenov, Synaptic inhibition controls transient oscillatory
  synchronization in a model of the insect olfactory system, Frontiers in
  Neuroengineering 5.

\bibitem{borgers2003synchronization}
C.~B{\"o}rgers, N.~Kopell, Synchronization in networks of excitatory and
  inhibitory neurons with sparse, random connectivity, Neural computation
  15~(3) (2003) 509--538.

\bibitem{catsigeras2010chaos}
E.~Catsigeras, Chaos and stability in a model of inhibitory neuronal network,
  International Journal of Bifurcation and Chaos 20~(02) (2010) 349--360.

\bibitem{buzsaki2004}
G.~Buzs{\'a}ki, A.~Draguhn, Neuronal oscillations in cortical networks, science
  304~(5679) (2004) 1926--1929.

\bibitem{wang1996gamma}
X.-J. Wang, G.~Buzs{\'a}ki, Gamma oscillation by synaptic inhibition in a
  hippocampal interneuronal network model, Journal of Neuroscience 16~(20)
  (1996) 6402--6413.

\bibitem{sohal2005inhibitory}
V.~S. Sohal, J.~R. Huguenard, Inhibitory coupling specifically generates
  emergent gamma oscillations in diverse cell types, Proceedings of the
  National Academy of Sciences 102~(51) (2005) 18638--18643.

\bibitem{white2000networks}
J.~A. White, M.~I. Banks, R.~A. Pearce, N.~J. Kopell, Networks of interneurons
  with fast and slow $\gamma$-aminobutyric acid type a (gabaa) kinetics provide
  substrate for mixed gamma-theta rhythm, Proceedings of the National Academy
  of Sciences 97~(14) (2000) 8128--8133.

\bibitem{echeveste2016drifting}
R.~Echeveste, C.~Gros, Drifting states and synchronization induced chaos in
  autonomous networks of excitable neurons, Frontiers in Computational
  Neuroscience 10.

\bibitem{rubin2008mmo}
J.~Rubin, M.~Wechselberger, The selection of mixed-mode oscillations in a
  hodgkin-huxley model with multiple timescales, Chaos: An Interdisciplinary
  Journal of Nonlinear Science 18~(1) (2008) 015105.

\bibitem{kispersky2010EHC}
T.~Kispersky, J.~A. White, H.~G. Rotstein, The mechanism of abrupt transition
  between theta and hyper-excitable spiking activity in medial entorhinal
  cortex layer ii stellate cells, PloS one 5~(11) (2010) e13697.

\bibitem{yamashita2008emergence}
Y.~Yamashita, J.~Tani, Emergence of functional hierarchy in a multiple
  timescale neural network model: a humanoid robot experiment, PLoS
  Computational Biology 4~(11) (2008) e1000220.

\bibitem{papo2013time}
D.~Papo, Time scales in cognitive neuroscience, Frontiers in Physiology 4.

\bibitem{siebenhuhner2016cross}
F.~Siebenh{\"u}hner, S.~H. Wang, J.~M. Palva, S.~Palva, Cross-frequency
  synchronization connects networks of fast and slow oscillations during visual
  working memory maintenance, Elife 5 (2016) e13451.

\bibitem{penn2016network}
Y.~Penn, M.~Segal, E.~Moses, Network synchronization in hippocampal neurons,
  Proceedings of the National Academy of Sciences 113~(12) (2016) 3341--3346.

\bibitem{gupta2016suppression}
K.~Gupta, G.~Ambika, Suppression of dynamics and frequency synchronization in
  coupled slow and fast dynamical systems, The European Physical Journal B
  89~(6) (2016) 147.

\bibitem{abbott2000synaptic}
L.~F. Abbott, S.~B. Nelson, Synaptic plasticity: taming the beast, Nature
  Neuroscience 3~(11s) (2000) 1178.

\bibitem{borst1999information}
A.~Borst, F.~E. Theunissen, Information theory and neural coding, Nature
  Neuroscience 2~(11) (1999) 947--957.

\bibitem{averbeck2004coding}
B.~B. Averbeck, D.~Lee, Coding and transmission of information by neural
  ensembles, Trends in Neurosciences 27~(4) (2004) 225--230.

\bibitem{stein1967information}
R.~B. Stein, The information capacity of nerve cells using a frequency code,
  Biophysical Journal 7~(6) (1967) 797--826.

\bibitem{izhikevich2003bursts}
E.~M. Izhikevich, N.~S. Desai, E.~C. Walcott, F.~C. Hoppensteadt, Bursts as a
  unit of neural information: selective communication via resonance, Trends in
  Neurosciences 26~(3) (2003) 161--167.

\bibitem{tiesinga2009cortical}
P.~Tiesinga, T.~J. Sejnowski, Cortical enlightenment: are attentional gamma
  oscillations driven by ing or ping?, Neuron 63~(6) (2009) 727--732.

\bibitem{womelsdorf2007attention}
T.~Womelsdorf, P.~Fries, The role of neuronal synchronization in selective
  attention, Current opinion in neurobiology 17~(2) (2007) 154--160.

\bibitem{stoop2004phase}
R.~Stoop, J.~Buchli, M.~Christen, Phase and frequency locking in detailed
  neuron models, in: Proceedings of the international symposium on nonlinear
  theory and its applications (NOLTA), 2004, pp. 43--6.

\bibitem{lowet2015input}
E.~Lowet, M.~Roberts, A.~Hadjipapas, A.~Peter, J.~van~der Eerden, P.~De~Weerd,
  Input-dependent frequency modulation of cortical gamma oscillations shapes
  spatial synchronization and enables phase coding, PLoS Computational Biology
  11~(2) (2015) e1004072.

\bibitem{chow1998frequency}
C.~C. Chow, J.~A. White, J.~Ritt, N.~Kopell, Frequency control in synchronized
  networks of inhibitory neurons, Journal of Computational Neuroscience 5~(4)
  (1998) 407--420.

\end{thebibliography}


\end{document}